\documentclass[pra,aps,twocolumn]{revtex4}
\usepackage{epsfig}

\begin{document}
\title{Efficient
upconversion of sub-THz radiation in a high-Q whispering gallery resonator}

\author{D. V. Strekalov}
\author{A. A. Savchenkov}
\author{A. B. Matsko}
\author{N. Yu}

\affiliation{Jet Propulsion Laboratory, California Institute of
Technology, 4800 Oak Grove Drive, Pasadena, California 91109-8099
}

\date{\today}

\begin{abstract}
We demonstrate efficient upconversion of sub-THz radiation into the optical domain in a high-Q whispering gallery mode resonator with quadratic optical nonlinearity. The $5\cdot 10^{-3}$ power conversion efficiency of continuous wave 100~GHz signal is achieved with only 16 mW of optical pump. \\
%$OCIS: 190.4223,\;190.4360,\;350.4010,\;040.2235 $
\end{abstract}

\maketitle

Detection of weak sub-THz signals, and ultimately counting of photons, is desirable for quantum communications, astronomy and spectroscopy applications. Photon counting in this range is achieved only at sub-Kelvin temperatures
\cite{komiyama2000,hashiba06,ikushima06,karasik07}. 
The origin of this difficulty can be
seen from the photon counting criterion introduced as a requirement that the photon energy $\hbar\omega$ must exceed the energy delivered by the minimal detectable signal power $P_{min}$ during the sampling time $\tau$:
\begin{equation}
P_{min}\tau<\hbar\omega.\label{counting}
\end{equation}
In most detectors, the minimal detectable power  $P_{min}$ is equal to the power of the noise, and therefore is called the noise equivalent power (NEP). If the noise power density is $S(\omega)$ and the frequency range of detectable signal is $\Delta\omega$,  relation (\ref{counting}) for a 100\%-efficient detector becomes
\begin{equation}
S(\omega)\Delta\omega\tau <\hbar\omega.\label{counting1}
\end{equation}
The physical meaning of the counting criterion in form (\ref{counting1}) is that a detector has to be fast enough to at least count the photons at the level of its own dark noise.
 The noise spectrum is ultimately limited by the
thermal fluctuations: $S(\omega)=k_BT$. This is a direct consequence of the equipartition theorem; also see theoretical discussion in \cite{Matsko08THzTheory}. If then the time
resolution of the detector $\tau$ is equal to the $1/\Delta\omega$, then the photon counting criterion simply reads $\hbar\omega>k_BT$.

The time resolution $\tau$ of the available sub-THz detectors \cite{komiyama2000,hashiba06,ikushima06,karasik07} is determined by the speed
of the transient processes and is independent of the detector frequency
range $\Delta\omega$. Typically $\tau\Delta\omega\gg 1$, which is one reason why they have to
operate at temperatures much lower than a single-photon
energy (1THz $\approx$ 48~K). Other reasons have to do with the
detector properties, such as the $T_c$ of the superconductors.
Naturally, it is desirable to create sub-THz photon counters operating at room temperature. A nonlinear conversion of low frequency photons into optical domain
\cite{chiou72apl,abbas76ao,albota04ol,temporao06ol,vandevender07josab,khan07} is useful here.
This approach should allow for $\tau\Delta\omega\ll 1$, because the photon-counting detector response time $\tau$ and the conversion frequency range $\Delta\omega$ are now determined by different physical processes.

Sub-THz photon counting based on nonlinear conversion has not been realized yet. The main limitation here is low conversion efficiency. The best known to us peak efficiency of power
conversion of approximately $6\cdot 10^{-4}$ was demonstrated in the 0.5-0.7 THz
range \cite{khan07} with pulsed optical pump of 500~W peak and
one watt average power \cite{khan07comment}. Therefore the CW power conversion efficiency was $1.2\cdot 10^{-6}$, which corresponds to the photon-number conversion efficiency of $3.7\cdot 10^{-9}$ as can be found from the Manley-Rowe relation.

We report on a three orders of magnitude improvement in the conversion efficiency, using a high-Q optical whispering gallery mode (WGM) resonator. WGM resonators have already been successfully used in nonlinear optics in general and in the  microwave photonics in particular
\cite{cohen01el-a,rabiei02jlt,ilchenko03mod,hosseinzadeh06mtt}.
Our achievement is in using this technique for efficient conversion of 100~GHz radiation. 

\begin{figure}[htp]
%\vspace*{-1in}
\centerline{
\input epsf
\setlength{\epsfxsize}{3.6in} \epsffile{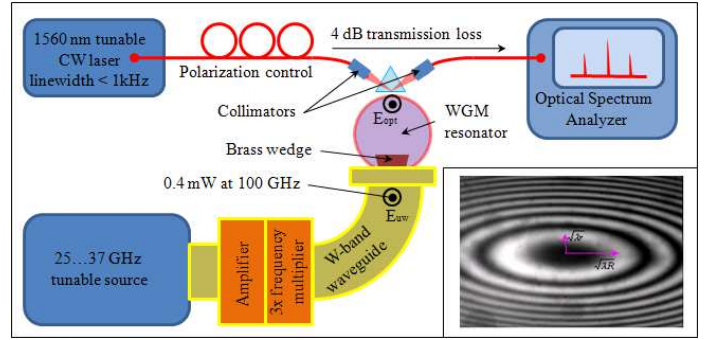} 
}
\caption[]{\label{fig:schematic}Schematic of the experiment setup. The WGM resonator is
coupled to the input and output free-space optical beams by a diamond prism, and RF-coupled to the signal field supplied by the waveguide. On the inset: a photograph of interference fringes arising between the rim of the resonator disk and a flat surface.}
\end{figure}

Our setup is shown in Fig.~\ref{fig:schematic}. The WGM resonator
disk made from a z-cut lithium niobate wafer has 1.8 mm radius
and 0.22 mm thickness. It is mounted on a brass support (not shown in Fig.~\ref{fig:schematic}) attached to the opening of a metal waveguide supplying the microwave signal. A
brass wedge inserted into the waveguide opening is used to optimize the signal field coupling into the disk. Both microwave and optical fields are polarized along the z-axis, as shown in the Figure, so that the largest nonlinearity component $r_{33}$ is employed.

The pump light of the wavelength $\lambda = 1560$ nm is coupled to the resonator using frustrated total internal reflection in a diamond prism. The coupling is adjusted by a few degree temperature variation of the thoroughly thermo-stabilized brass support. This changes the gap between the
resonator and the prism, due to the
differential thermal expansion of the materials. Sub-degree variations of the
temperature are used to tune the WGMs frequencies, due to thermal
expansion of the disk.

When the incident light reflects from the back side of the prism, its evanescent field propagates outside along the prism surface with the wave vector projection $k_{||}=n_d(\omega/c)\sin\theta$,
where $n_d$ is the refraction index of diamond and $\theta$ is the incidence angle. The wave vectors of light inside and outside of the disk should match, which yields the following condition for the incidence angle:
\begin{equation}
\sin\theta=n_e/n_d.\label{k2}
\end{equation}
In (\ref{k2}), $n_e$ is the extraordinary refraction index of lithium niobate. To establish the second condition of efficient coupling, we visualize the coupling region as the first-order Newton ``ring" arising between the resonator disk rim and a flat surface, see the inset of Fig.~\ref{fig:schematic}. The ratio of the ellipse axes is equal to the square root of the ratio of the two local curvature radii $R$ (the disk radius) and $r$ of the rim. It is congruent with the incident beam ``footprint" if
\begin{equation}
r/R=(\cos\theta)^2.\label{aspect}
\end{equation}
Given $\theta$, condition (\ref{aspect}) yields the optimal $r/R$ ratio.

\begin{figure}[htp]
%\vspace*{-1in}
\centerline{
\input epsf
\setlength{\epsfxsize}{3in} \epsffile{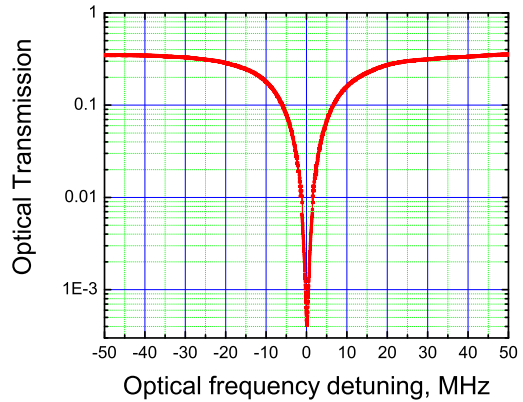} }
\caption[]{\label{fig:goodcoupling} A high-contrast optical WGM resonance.
}
\end{figure}

A WGM resonance can be observed as absorption of the light transmitted from the laser source to the spectrum analyzer (see Fig.~\ref{fig:schematic}). In our experiment the contrast of the resonance reached 99.96\%, see Fig.~\ref{fig:goodcoupling}. Notice that the baseline in Fig.~\ref{fig:goodcoupling} is below the unity. This 4 dB broadband loss originates from the reflections of two diamond surfaces, the fiber tips and collimation lenses, as well as from the possible mismatch of the collection optics. The latter suggests that the observed high contrast could be partially contributed to by some interference phenomena, e.g. the fringes in the output light beam. Although we made an effort to minimize these effects by a thorough alignment procedure, an additional study of the WGM resonators coupling is required.

\begin{figure}[htp]
%\vspace*{-1in}
\centerline{
\input epsf
\setlength{\epsfxsize}{3in} \epsffile{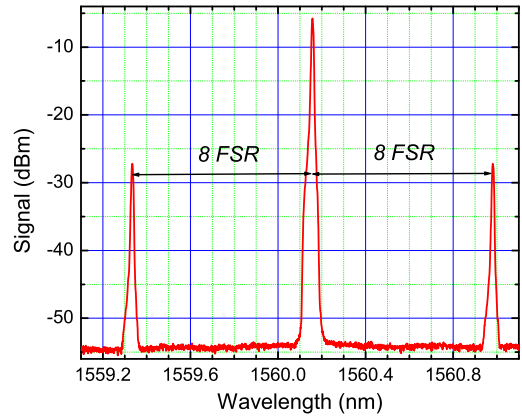} }
\caption[]{\label{fig:spectrum}The spectrum of the optical output in presence of the microwave signal. The
signal power at the waveguide input is approximately 0.4 mW. The optical power at the coupling prism input is 16 mW.}
\end{figure}

First we observed the optical WGM spectrum of our resonator by
sweeping the laser frequency, and measured the free spectral range (FSR) $\Omega=2\pi\cdot 12.64$~GHz and $\Delta\omega\approx 2\pi\cdot 20$~MHz, which yields
the quality factor $Q\approx 10^7$. Then we set the
laser at one of the WGMs and varied the signal
frequency around $\omega_{rf}=8\Omega\approx2\pi\cdot 101.12$ GHz. The sidebands
were observed at 101.38 GHz (see Fig.~\ref{fig:spectrum}). It is
interesting to point out that in this case both Stokes and
anti-Stokes sidebands have equal amplitudes; if however we change
the signal frequency we can have one sideband exceed the other
or vice versa, depending on the sign of the detuning. This can be considered as an
evidence of significant WGM frequency dispersion on the span of 16 FSRs.

The employed method of microwave coupling into the disk is rather
inefficient, since the main part of the microwave energy
likely reflects back into the waveguide or scatters into space.
However the CW power conversion efficiency into \emph{each} Stokes
and anti-Stokes sideband was found to be $5\cdot 10^{-3}$, which
corresponds to a photon-number conversion efficiency of $2.6\cdot
10^{-6}$ into each sideband, and $5.2\cdot
10^{-6}$ into both bands. This greatly improves the results of \cite{khan07}.

To estimate how closely our experiment comes to the photon counting limit (\ref{counting1}) we find the experimental value of the NEP density $S(\omega)$ and subsitute into (\ref{counting1}), finding the maximum detection bandwidth $\Delta\omega_{max}$ allowing for photon counting. The resolution bandwidth of the optical spectrum analyzer was 1.23 GHz. Each Stokes and anti-Stokes signal observed within this bandwidth was some 27 dB above the instrumental noise floor ( see Fig.~\ref{fig:spectrum}), while the input microwave power was 0.4 mW. Therefore the NEP density can be found as the power-to-bandwidth ratio less 27 dB, which yields $S(\omega)=1.6$ fW/Hz and $\Delta\omega_{max} = 1.3$ Hz. Notice that if we increase the conversion efficiency to unity, this bandwidth will increase to 0.52 MHz, which would require $Q\approx 4\cdot 10^{8}$. This quality factor value is certainly within reach \cite{savchenkov07ol}.

A unity conversion efficiency in continuous regime is theoretically predicted \cite{Matsko08THzTheory}
for the all-resonant up-converters, supporting the microwave WGM as well as the optical ones. The interacting modes have to satisfy the phase-matching conditions $\omega_{rf}=L_{rf}\Omega$ and $L_{rf}=L_{a}-L_p$,
where $L_{rf}, \,L_{a}$ and $L_p$ are the orbital momenta of the
THz, anti-Stokes and pump WGMs, respectively, and $\Omega$ is the
optical FSR of the resonator. These conditions can be satisfied by
taking advantage of the strong waveguide dispersion of the sub-THz mode.

Unity conversion efficiency can be achieved when the loss rate of the sub-THz
mode $\gamma=\gamma_{nl}+\gamma_{abs}$ is dominated by the rate of
nonlinear frequency conversion: $\gamma_{nl}\gg\gamma_{abs}$. Furthermore, the optical WGMs need to be strongly coupled to the free space, so that for the optical loss rates $\Gamma >\Gamma_{abs}$ would hold. The detection bandwidth $\Delta\omega$ is then equal to that of the loaded optical WGM.

The theoretical value found in
\cite{Matsko08THzTheory} for the NEP spectral density is $S=2k_BT$, for a resonator which is in thermal equilibrium with the surrounding heat bath. If the
resonator is decoupled from the bath, its own thermal fluctuations
contribution to the microwave noise is suppressed by virtue of the
fluctuation-dissipation theorem. Then instead of $T$ one should use
$T_{eff}=T\Gamma_{abs}/\Gamma<T$, which can bring the noise level
below the classical limit.

For a conservative estimate we assume that the resonator is
coupled with a thermal bath at $T=300$ K. We further assume $Q=10^8$
\cite{savchenkov07ol}, $\lambda=1560$ nm (so that $\Delta\omega\approx2\pi\cdot 2$ MHz), and $\tau = 5$ ns. Then from
the photon counting criterion (\ref{counting1}) we find
$\omega >2\pi\cdot 0.12$ THz, which means that counting 0.12~THz photons should be possible. If we decouple the resonator from the thermal bath it should
be possible to count photons at even lower frequencies,
and to reach higher sensitivity.

To summarize, we have demonstrated nonlinear conversion of sub-THz radiation to optics with efficiency greatly surpassing the state of the art. The NEP density in our measurement was a factor of $5\cdot 10^{-6}$ worse than the theoretical limit $2k_BT =8\cdot 10^{-21}$ W/Hz for an all-resonant WGM converter with unity efficiency \cite{Matsko08THzTheory}. This factor is very close to the two-band photon-number conversion efficiency $5.2\cdot 10^{-6}$ we have measured, which means that the deficiency of our experiment is solely due to insufficient conversion efficiency.
We believe that realization of this method with an all-resonant WGM converter has a potential for achieving sub-THz photon counting at room temperature. 
Benefits from its practical implementation are expected in the
areas of quantum information and computing (especially based on
quantum electronic circuits), astronomy, and spectroscopy.

The research described in this paper was carried out at the Jet
Propulsion Laboratory, California Institute of Technology, under a
contract with the NASA.


\begin{thebibliography}{99}
\bibitem{khan07comment} In \cite{khan07} the actual measured efficiency which we quote here is multiplied by approximately a factor of five to account for the loss due to optical components.
\end{thebibliography}
\end{document}